\documentclass[10pt, conference, letter]{IEEEtran}

\usepackage{amssymb}
\usepackage{amsfonts}
\usepackage{epsf}
\usepackage{epsfig}
\usepackage{epstopdf}
\usepackage{array}
\usepackage{amsmath}
\usepackage{url}
\usepackage[algoruled, linesnumbered]{algorithm2e}

\usepackage{amsthm}
\usepackage{xifthen}
\usepackage{mathrsfs}
\usepackage[mathscr]{euscript}
\usepackage[sans]{dsfont}
\usepackage[dvipsnames]{xcolor}
\usepackage{graphicx}
\usepackage{caption}
\usepackage{subcaption}

\usepackage{epsf}
\usepackage{epsfig}

\usepackage{array}

\usepackage{times}
\usepackage{graphicx}
\usepackage{amsmath}
\usepackage{amssymb}
\usepackage{amsxtra}
\usepackage{here}
\usepackage{rawfonts}
\usepackage{times}
\usepackage{url}
\usepackage{cite}
\usepackage{amsthm}
\usepackage{graphicx}

\usepackage{cite} 

\usepackage[T1]{fontenc}
\usepackage[utf8]{inputenc}
\usepackage{authblk}
\IEEEoverridecommandlockouts

\theoremstyle{definition}

\ifCLASSINFOpdf

\else

\fi

\newlength{\aligntop}
\newlength{\alignbot}
\makeatletter

\begin{document}

\title{QoS and Jamming-Aware Wireless Networking Using Deep Reinforcement Learning  \vspace{-0.7ex} 
}

\author[D. Kopta et al.]
	{\normalsize Nof Abuzainab$^*$, Tugba Erpek$^*$, Kemal Davaslioglu$^*$, Yalin E. Sagduyu$^*$, Yi Shi$^*$, Sharon J. Mackey$^{\dagger}$, Mitesh Patel$^{\dagger}$,  \\  Frank Panettieri$^{\dagger}$, Muhammad A. Qureshi$^{\dagger}$, Volkan Isler$^\ddagger$, and Aylin Yener$^{\S}$
		\\
		$^*$Intelligent Automation Inc., Rockville, MD 20855,  USA,$^{\dagger}$U.S. Army C5ISR, APG, MD 21005, USA,\\
 $^\ddagger$University of Minnesota, Minneapolis, MN 55455 USA,$^{\S}$The Pennsylvania State University, University Park, PA 16802, USA\\
		\thanks{DISTRIBUTION A. Approved for public release: distribution unlimited.}
		\thanks{\textsuperscript{\textcopyright} 2019 IEEE. Personal use of this material is permitted. Permission from IEEE must be obtained for all other uses, in any current or future media, including reprinting/republishing this material for advertising or promotional purposes, creating new collective works, for resale or redistribution to servers or lists, or reuse of any copyrighted component of this work in other works.}
		\vspace{-3ex} 

}

\maketitle

\begin{abstract}
The problem of quality of service (QoS) and jamming-aware communications is considered in an adversarial wireless network subject to external eavesdropping and jamming attacks. To ensure robust communication against jamming, an interference-aware routing protocol is developed that allows nodes to avoid communication holes created by jamming attacks. Then, a distributed cooperation framework, based on deep reinforcement learning, is proposed that allows nodes to assess network conditions and make deep learning-driven, distributed, and real-time decisions on whether to participate in data communications, defend the network against jamming and eavesdropping attacks, or jam other transmissions. The objective is to maximize the network performance that incorporates throughput, energy efficiency, delay, and security metrics. Simulation results show that the proposed jamming-aware routing approach is robust against jamming and when throughput is prioritized, the proposed deep reinforcement learning approach can achieve significant (measured as three-fold) increase in throughput, compared to a benchmark policy with fixed roles assigned to nodes.
\end{abstract}

%\begin{IEEEkeywords}
%component, formatting, style, styling, insert
%\end{IEEEkeywords}

\section{Introduction}
Congested and contested environments in wireless networks are characterized by intermittent connectivity due to severe interference conditions including jamming. In addition, adversaries may attempt to eavesdrop on the ongoing transmissions as a stealth attack. Wireless nodes are typically limited by different levels of resources such as energy (e.g., battery), computation (e.g., processing power and memory), and communication resources (e.g., bandwidth). Therefore, effective coordination and cooperation among network nodes is necessary to manage information flows and satisfy the requirements of communications subject to adversarial effects (jamming and eavesdropping) and resource limitations.

The problem of secure routing with quality of service (QoS) considerations has attracted recent attention in adversarial network settings. 
A wireless communication network in the presence of eavesdroppers is considered in \cite{secQoS1} and a centralized approach is proposed to construct a routing path that maximizes the system performance, in terms of the secrecy outage probability.  
An ant colony optimization algorithm is proposed in \cite{secQoS2} to determine the optimal routing path that ensures the desired security and QoS requirements. These solutions are centralized and do not necessarily respond effectively to changes in the network.

To adapt to network dynamics, reinforcement learning has emerged as a viable solution for distributed network optimization. Reinforcement learning allows nodes acting as agents to select their decisions based on the rewards received from the environment and to adapt their decisions in real time to the changes in the environment. Reinforcement learning has been applied to develop distributed resource allocation and routing algorithms in wireless networks \cite{RL1,RL2,RL3,RL4}. The main drawback of conventional reinforcement learning algorithms is that the computational overhead involving maintenance of the Q-table grows significantly with the state and action space \cite{Arul}, which limits their practicality.

More recently, deep reinforcement learning has been proposed to reduce the computational burden of conventional reinforcement learning, leading to a more practical implementation. Deep reinforcement learning relies on deep neural networks to determine the optimal decisions for a given state of the environment, and thus, its computational complexity is less compared to conventional reinforcement learning. A QoS-aware routing protocol based on deep reinforcement learning is proposed in \cite{DRL1} that determines, based on the traffic demands, the routing paths that maximize the number of flows with satisfied QoS requirements. Jamming is a major security concern for performance loss in wireless networks \cite{jamming}, and deep learning has been shown to be effective in both launching and defending against jamming attacks \cite{Erpek, Davaslioglu}. An anti-jamming policy using deep reinforcement learning is developed in \cite{DRL2} that exploits mobility and spread spectrum, and allows a mobile user to escape an area of heavy jamming. 
The underlying reinforcement learning algorithm determines a global optimal path. Hence, the decision is not individually performed by nodes in a distributed fashion. This raises the need to design a distributed solution that achieves the desired network performance and is simultaneously robust to both jamming and eavesdropping attacks.

In this paper, we present a distributed adaptive framework for cooperation and coordination in an adversarial wireless network environment that consists of blue-force and red-force nodes. Network communications and security tasks share resources and it is beneficial to design and perform these tasks jointly \cite{comm_EW}. We assume that blue-force nodes aim to maximize their throughput in the presence of jamming from red-force nodes, prevent red-force nodes from eavesdropping their transmissions, or jam red-force transmissions when they detect red-force signals. The contributions of the paper are summarized below:
\begin{itemize}
\item We consider a novel distributed approach that allows nodes to assess the current network conditions in terms of the received signal strength, queue length, location, and messages, dynamically adjust their roles, and decide whether to participate in communication or defend the network against attacks or attack other (out-of-network) nodes.
\item We design distance-vector based interference-aware distributed routing protocols (one jamming-avoiding, and the other jamming-aware) that allow nodes to construct a routing path to avoid communication holes due to jamming. The proposed routing protocols rely at most on two-hop information to make routing decisions. Thus, they do not incur significant communication overhead.
\item We develop a deep reinforcement learning solution for nodes to decide on whether to participate in communication, defend the network, or attack other transmissions for the sake of optimizing the desired performance metric, which jointly captures throughput, energy efficiency, delay, and security aspects. Nodes decide on one of five roles, namely transmitter, receiver, cooperative jammer (jamming red-force eavesdropping nodes so that they cannot decode blue-force transmissions), adversarial jammer (jamming blue-force data transmissions), or waiting.
\item Simulation results demonstrate the robustness of the proposed jamming-aware routing protocol and show that our proposed deep reinforcement learning approach achieves a three-fold increase in throughput,  compared to a benchmark policy that assumes fixed roles for nodes.
\end{itemize}

The rest of the paper is organized as follows. Section \ref{Sec:Sec2} describes the system model. Section \ref{Sec:Sec3} presents deep reinforcement learning approach. Section \ref{Sec:Sec4} reports the simulation results. Section \ref{Sec:Sec5} concludes the paper. 

\section{System Model} \label{Sec:Sec2}
Consider a wireless network of $N$ blue-force nodes, $\mathcal{N}$, that are uniformly deployed over a disk of radius $R$ and move according to the Brownian motion mobility model \cite{Brownian}. Nodes communicate with each other using carrier sense multiple access/collision avoidance (CSMA/CA) medium access protocol. In the network, $\mathcal{F}$ communication flows are present and the pair $(s_i,d_i)$ corresponds to the source/destination pair of the $i$th communication flow. To deliver packets from each source to its destination, link state geometric routing is employed in which each node chooses, among its neighbors (nodes within its communication range $r$), the next hop node with the minimum distance to the destination. Distance will be defined later in this section. Each node maintains a queue for the  packets received from each communication flow and a routing table that stores the next hop for each flow and the minimum distance to its destination. Thus, given the presence of packets pertaining to multiple flows, each node $i$ selects, at any time, the flow and the next hop to minimize the distance to the destination $D$, i.e.,
\begin{equation}
j^*=\arg\min_{j \in \mathcal{N}_i} d_{ij} + d_{j_D},
\end{equation}
where $d_{ij}$ is the distance between node $i$ and $j$ and $\mathcal{N}_i$ is the set of neighbors of node $i$.

Red-force nodes are uniformly distributed at the boundary of the network and seek to attack the communication of blue-force nodes. In particular, among red-force nodes, $M_{CJ}$ nodes choose to eavesdrop communications of blue-force nodes, $M_{AJ}$ nodes perform adversarial jamming, and $M_T$ nodes are engaged in their own communications. Due to jamming, communication holes occur in the network, which will considerably reduce the network throughput through its impact on network-level communications. At the network level, when the next hop is jammed in conventional geometric routing, the transmitter is not aware of the status of the next hop and keeps sending packets indefinitely to the same next hop. As a consequence, packets will never arrive to its destination, and throughput will significantly be degraded. We consider two interference-aware enhancements to the conventional geometric routing in the following  subsection.

\subsection{Enhanced Geometric Routing Protocols}
 The first enhanced routing protocol is the \emph{jamming-avoiding geometric routing} in which a jammed node broadcasts its status to its immediate neighbors. To detect adversarial jamming,  blue-force  nodes sense the spectrum. If the received signal strength is higher than a certain threshold, the node considers itself jammed. Then its neighbors set the distance of the jammed node to the destinations to infinity, and the routing tables are updated accordingly. The second enhanced protocol, termed the \emph{adaptive, jamming-aware geometric routing protocol}, aims at constructing the minimum distance path to the destination such that it avoids the jammed area. This routing protocol requires two-hop network information, namely each transmitter $i$ obtains information not only about its neighbors but the neighbors of its neighbors. This information includes the jamming status of each node and its distance to its destination. Transmitter $i$ then computes, for each unjammed neighbor $j$, the angle $\theta_{ijj'}$ between $i$, $j$, and the next hop $j'$ from $j$. The cost is the Euclidean distance between transmitter $i$ and next node $j$ divided by  $\theta_{ijj'}$. Then transmitter $i$ chooses as the next hop, the unjammed neighbor $j^*$ with the minimum cost, i.e.,
\begin{equation}
j^*=\arg\min_{j \in \mathcal{N}_i}\frac{ d_{ij}}{\theta_{ijj'}} + d_{j_D}.
\end{equation}

\begin{figure}[t]{
	\centering 
	\includegraphics[width=8 cm,height=3.25cm,angle=0]{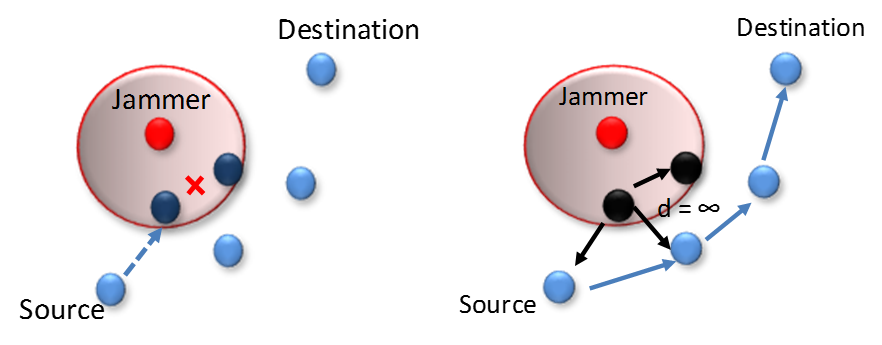}
	\caption{ Left – minimum (geographic) distance routing, right – jamming-avoiding geometric routing.}\label{routing1}
	}
\end{figure}

\begin{figure}[t]{
	\centering
	\includegraphics[width=4 cm,height=4cm,angle=0]{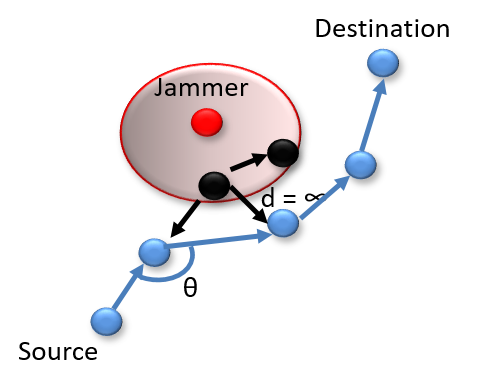}
	\caption{Adaptive, jamming-aware geometric routing.}\label{routing2}
	}
\end{figure}

Figures \ref{routing1} and \ref{routing2} illustrate the two routing protocols.


\subsection{Role Selection for QoS-aware and Secure Communications}
To ensure successful packet delivery while keeping transmissions as secure as possible from red-force nodes,  each blue-force node uses local information such as received signal strength, neighbor information, queue length, and distance to destination, and chooses at each time instant to perform one of the following actions: 1) transmit a packet,  2) receive a packet, 3) cooperative jamming to mitigate eavesdropping attacks \cite{Yener}, 4) adversarial jamming to jam red-force nodes, or 5) wait. The objective is to determine the optimal action of nodes that maximizes the network utility within duration $T$. Nodes choose their actions according to $\boldsymbol{a}(t)=(a_i(t))_{i \in \mathcal{N}}$, where $a_i(t)$ corresponds to the action taken by node $i$ at time $t$. Network utility consists of three rewards and two costs. Reward $N_T(\boldsymbol{a}(t))$ for throughput is the number of successfully delivered packets at time $t$, reward $N_{CJ}(\boldsymbol{a}(t))$ for cooperative jamming is the expected number of failed eavesdropping attempts normalized by the total number of eavesdropping attempts at time $t$, and reward $N_{AJ}(\boldsymbol{a}(t))$ for adversarial jamming is the expected proportion of jammed red-force nodes at time $t$. At time $t$, cost $N_E(\boldsymbol{a}(t))$ is the energy consumed and cost $N_D(\boldsymbol{a}(t))$ for delay is the expectation of time when a node with non-empty queue decides to wait. The average over time for each metric is given by $\bar{N}_k (\boldsymbol{[a]}_1^T)=\frac{1}{T}\sum_{t=1}^T N_k(\boldsymbol{a}(t))$, where $k \in \{T, CJ,AJ, D, E\}$ and $\boldsymbol{[a]}_1^T =  (\boldsymbol{a}(1), \boldsymbol{a}(2),...,\boldsymbol{a}(T))$.

The overall optimization problem is given by
\begin{equation}
\max_{\boldsymbol{[a]}_1^T} \sum_{k \in \{ \mathcal{K}_R  \} } w_k \bar{N}_k (\boldsymbol{[a]}_1^T) - \sum_{k \in \{ \mathcal{K}_C  \}} w_k \bar{N}_k (\boldsymbol{[a]}_1^T), \label{eq:6}
\end{equation}
where reward types are $\mathcal{K}_R = \{T, CJ, AJ\}$, cost types are  $\mathcal{K}_C = \{E, D\}$, and weight $w_k$ is the respective weight of the individual metric $\bar{N}_k$.  
To solve (\ref{eq:6}), distributed reinforcement learning is employed next.
\section{ Deep Reinforcement Learning for QoS-aware and Secure Communications} \label{Sec:Sec3}
%\subsection{Reinforcement Q-learning}
\subsection{States, Actions, Rewards, and Costs}
To solve (\ref{eq:6}) with reinforcement learning, each node observes its state $\boldsymbol{s}_i(t)$  based on the messages received from its neighbors, performs an action $a_i(t)$, and receives a reward in return. The state $\boldsymbol{s}_i(t)$ of each node $i$ is  $\boldsymbol{s}_i(t)=(s_i^1(t),s_i^2(t), s_i^3(t), s_i^4(t), s_i^5(t), s_i^6(t), s_i^7(t))$, where
\begin{itemize}
\item $s_i^1(t)$: Node $i$'s queue length.
\item $s_i^2(t)$: The indicator if ready to send (RTS) message intended to node $i$  is received.
\item$s_i^3(t)$: The indicator if RTS not intended to node $i$ is received.
\item$s_i^4(t)$: The indicator if  clear to send (CTS) message intended to node $i$ is received.
\item $s_i^5(t)$: The indicator if CTS not intended to node $i$  is received.
\item $s_i^6(t)$: The indicator if node $i$ sent RTS.
\item $s_i^7(t)$: The indicator if node $i$ sent CTS.
\item $s_i^8(t)$: Number of time slots elapsed since an RTS is sent.
\item$s_i^9(t)$: Number of time slots elapsed for backoff, and
\item$s_i^{10}(t)$: Number of time slots elapsed waiting for data.
\end{itemize}

The action taken by each node is either Transmit (T), Receive (R), Cooperative Jamming (CJ), Adversarial Jamming (AJ), or Wait (W). Rewards associated for each action $a_i(t)$ given state $\boldsymbol{s}_i(t)$ are described below. 

\textbf{Transmit (T)}: When node $i$ decides to transmit, the reward is
 
\begin{equation}
r_i (a_i(t),\boldsymbol{s}_i(t)) = I(\textrm{SINR} > \tau) \times w_T \times (1- I(\textrm{CTS}(j,k))),
\end{equation}
  where $I(\textrm{SINR} > \tau)$ is the indicator that the signal-to-interference-plus-noise ratio (SINR) is greater than the required threshold $\tau$, which also indicates a successful packet delivery; $w_T$ is the weighing factor of throughput; and  $I(\textrm{CTS}(j,k))$ is the indicator that node $i$ received CTS which is not intended to it  (CTS sent from node $j$ to $k$), which indicates a  transmission collision. 
     
The cost of transmitting is the energy spent per time slot. In the considered scenario, all nodes transmit with fixed power. Therefore, one unit of energy is consumed by each node in each time slot, and the cost is given by $c_i(a_i(t),\boldsymbol{s}_i(t)) =  w_E$, where $w_E$ is the weighing factor of energy consumption.

\textbf{Receive (R)}: When node $i$ decides to receive, the reward is
\begin{equation}
r_i (a_i(t),\boldsymbol{s}_i(t)) = I (\textrm{SINR}>\tau) \times w_T \times I(\textrm{RTS}(i,k)),
\end{equation}
where  $I(\textrm{RTS}(i,k))$ indicates that node $i$ received an RTS intended to itself. 

The cost of receiving is the delay incurred on the packets stored in the node’s queue. In particular, it is given by $c_i(a_i(t),\boldsymbol{s}_i(t)) = I(q>0) \times w_D$, where  $I(q>0)$ is the indicator that node $i$'s queue length $q$ is positive; and  $w_D$ is the weighing factor of delay.

\textbf{Cooperative jamming (CJ)}: When node $i$ chooses to be a cooperative jammer, the reward is given by the proportion of failed eavesdropping attempts. In particular, the reward is 
\begin{equation}
r_i(a_i(t),\boldsymbol{s}_i(t)) = N_E \times w_{CJ}  \times I(\textrm{RTS}(j,k)),
\end{equation}
where $N_E$ is the average proportion of eavesdroppers within the transmitter’s range and the cooperative jammer range, which is given by the length of arc formed by the intersection of the network border, the reception area, and the jammed area  divided by the length of arc formed by the intersection of the network border and the reception area (note that the transmission reception area is approximated by a disk centered at the transmitter for computational tractability); $w_{CJ}$ is the weighing factor of cooperative jamming; and	$I\textrm{(RTS}(j,k))$ is the indicator that node $i$ received an RTS not intended to it, which indicates that node $i$ sensed an existing blue-force transmission.

The cost of cooperative jamming is the energy spent due to jamming and the incurred delay on node $i$’s packets. Thus, the cost is given by 
\begin{equation}
c_i(a_i(t),\boldsymbol{s}_i(t)) = w_E + I(q>0) \times w_D.\label{FJ}
\end{equation}
\\ \textbf{Adversarial jamming (AJ)}: When node $i$ acts as an adversarial jammer, the reward is computed by the proportion of jammed red-force receivers. Therefore, the reward is given by
\begin{equation}
    r_i(a_i(t),\boldsymbol{s}_i(t)) = N_{AJ} \times w_{AJ}  \times I_{\textrm{adv}},
\end{equation} where $N_{AJ}$ is the average proportion of red-force receivers in the jamming range (since the red-force nodes are uniformly distributed over the network border, the proportion $N_{AJ}$ is given by the length of arc formed by the intersection of the network border and the jammed area, divided by $2\pi R_{\max}$, where $R_{\max}$ is the radius of the network area);	$w_{AJ}$ is the weighing factor of adversarial jamming; and $I_{\text{adv}}$ is the indicator of sensed red-force transmission.

The cost of adversarial jamming is the energy spent due to jamming and the incurred delay on node $i$’s packets. Thus, the cost is given by (\ref{FJ}).

\textbf{Wait (W)}: Node $i$ chooses to wait when there is no benefit of performing any of other actions. i.e., when their utilities are negative. Thus, the utility of  waiting is set to zero.

Reinforcement Q-learning is adopted by approximating the value $Q$ function for a given state and action based on the utility $p(\boldsymbol{s}_i(t), a_i(t))$ of the current time slot $t$ according to:\\
\vspace{-0.4 cm}
\begin{eqnarray}
&&\hspace{-0.7 cm}Q_{\text{new}}(\boldsymbol{s}_i(t), a_i(t))= (1-\alpha)Q(\boldsymbol{s}_i(t), a_i(t))+ \alpha(p(\boldsymbol{s}_i(t), a_i(t))\nonumber\\
&&\hspace{2.2 cm}+\gamma \max_b Q(\boldsymbol{s}_i(t+1),b),
\end{eqnarray}
where $\alpha$ is the  learning rate, $\gamma$ is the discount factor, and $p(\boldsymbol{s}_i(t), a_i(t))$ is the immediate utility. The pseudocode of the Q-learning algorithm is given by Algorithm 1. The drawback of conventional reinforcement learning approaches is that the magnitude of Q-value evaluations grows exponentially with the state and action space. Thus, we consider an alternative approach, using deep learning, to approximate the Q value.
\begin{algorithm}[t] 
\small
\SetAlgoLined
%\KwResult{Write here the result }
 Initialize action-value function $Q$\;
\For{t=1,...,T }{
 \While{at least one flow $l$ has not reached to its destination, }{

\For{all nodes $i \in \mathcal{N}$}
{
\begin{enumerate}
\item Observes its state and acquires states of
\\ other nodes from local messages to form 
\\ observation vector $\boldsymbol{s}_i(t)$
\item Selects action 
\\$a_i(t)=\arg\max_{a_i(t) \in \mathcal{A}} Q(\boldsymbol{s}_i(t), a_i(t))$ \\ with probability (w.p.) $1-\epsilon$ or a random\\ action $a_i(t) \in \mathcal{A}$  w.p. $\epsilon$
\item Executes an action $a_i(t)$
\item Receives reward $R_i(\boldsymbol{s}_i(t), a_i(t))$
\item Broadcasts its action $a_i(t)$ to its neighbors. 

\end{enumerate}

}

 }
}
 \caption{Q-learning algorithm}\label{Qlearning}
\end{algorithm}
\normalsize

\subsection{Deep Q-learning}
 A deep Q-Network (DQN) is implemented to teach each node which action to perform given observations \cite{DQN}. In particular the actor-critic DQN algorithm is implemented. In the actor-critic DQN implementation, there are two deep neural networks with the same architecture but different parameters. The first network is the actor network that determines the optimal policy for a given current state. The second network is the critic network that evaluates the goodness of the chosen policy by the actor. 
The input of both the actor and the critic networks is the current state. The critic network generates an approximate value that will be used to approximate the Q value for a given state and action. The output of the actor network is an approximate policy that is the probability distribution $\pi_\theta (\boldsymbol{s}_i(t),a_t )$ of the action values for a given state $\boldsymbol{s}_i(t)$.

Since the actor and critic networks are two different networks, two distinct set of weights will be used to train each network. We denote by $\theta$ and $w$, the set of weights for the actor and critic, respectively.
The critic network is trained (i.e., the weights are updated) such that the loss function $L(w)$ is minimized. $L(w)$ is the squared error between the target value and the approximate value generated by the critic network, i.e., 
\begin{eqnarray}
L(w)&=&(p(\boldsymbol{s}_i(t),a_t ) \nonumber \\  && \hspace{-0.3cm} + \gamma \max_a  Q_w (\boldsymbol{s}_i(t+1),a)-Q_w (\boldsymbol{s}_i(t),a(t)))^2, 
\end{eqnarray}
where $Q_w(s,a)$ is the approximate Q value generated by the critic network. The weights of the actor, on the other hand, are updated according to a policy gradient. Thus, the weights of the actor and the critic are updated, respectively as:
\begin{itemize}
\item Policy Update for actor:
\begin{equation}
\Delta \theta=\alpha \nabla_\theta\log \pi_\theta (\boldsymbol{s}_i(t),a(t)) Q_w (\boldsymbol{s}_i(t),a(t)).
\end{equation}
\item Value Update for critic:
\begin{eqnarray}
&&\hspace{-1.5 cm}\Delta w=(p(\boldsymbol{s}_i(t),a_t )+\gamma \max_a  Q_w (\boldsymbol{s}_i(t+1),a)\nonumber\\
&&-Q_w (\boldsymbol{s}_i(t),a))\nabla_w Q_w (\boldsymbol{s}_i(t),a(t)),
\end{eqnarray}
\end{itemize}
where $\alpha$ and $\beta$ are the learning rates of the actor and critic network, respectively.
To reduce the training time and increase the convergence, experience replay is used in which an agent stores its experience into a memory and pools over many episodes into a replay memory. This enables to reduce the effects of correlated training data. The replay memory buffer also helps improve stability by regularizing the policy. The pseudocode for the DQN algorithm is shown in Algorithm 2.

\begin{algorithm}[t]
\small
\SetAlgoLined
%\KwResult{Write here the result }
 Initialize replay memory $\mathcal{M}$ with capacity $C$\;
 Initialize action-value function $Q$  with random weights\;
\For{episode=1,...,M }{
 \While{at least one flow $l$ has not reached to its destination, }{

\For{all nodes $i \in \mathcal{N}$ (in parallel)}
{
\begin{enumerate}
\item Observes its state and acquires states of other
\\ nodes from local messages to form observation
\\ vector $\boldsymbol{s}_i(t)$
\item Selects action 
\\$a_i(t)=\arg\max_{a_i(t) \in \mathcal{A}} Q(\boldsymbol{s}_i(t), a_i(t))$ \\ w.p. $1-\epsilon$ or a random action $a_i(t) \in \mathcal{A}$ w.p. $\epsilon$
\item Executes an action $a_i(t)$
\item Receives reward $R_i(\boldsymbol{s}_i(t), a_i(t))$
\item Broadcasts its action $a_i(t)$ to its neighbors. 
\item Stores transition in replay memory $\mathcal{M}$
\item  Performs weight updates 
\end{enumerate}

}

 }
}
 \caption{Deep Q-learning algorithm}\label{DQN}
\end{algorithm}
\normalsize

\section{Simulation Results} \label{Sec:Sec4}
\subsection{Interference-aware routing results}
For network simulations written in Matlab, we consider the following network settings. Number of blue force nodes is 40, number of red-force nodes is 10 ($M_{CJ}$ is 3, $M_{AJ}$ is 3, and $M_T$ is 4), network radius $R$ is $10$ km, speed of Brownian motion is $1$ m/s. Transmit power of each node is $1$ W. Channel model follows free space propagation with log-normal shadowing. SINR threshold $\tau$ is $5$ dB. We simulated three routing protocols, namely minimum (geographic) distance routing, jamming-avoiding geometric routing and adaptive, jamming-aware geometric routing. Minimum distance routing is the baseline protocol that does not incorporate node jamming status as a routing metric; as a result, it is expected that it provides the worst performance results. Each simulation is run for $15,000$ time slots. To assess the robustness of the proposed routing protocols, jamming by the red-force nodes is activated after the first 5,000 time slots. The same scenario is run for all routing protocols by fixing the seed of random number generator in each experiment. The simulation is repeated $30$ times with different seeds to determine the confidence intervals on the results. 

\begin{figure}[t]{
	\centering
	\includegraphics[width=7 cm,height=4 cm,angle=0]{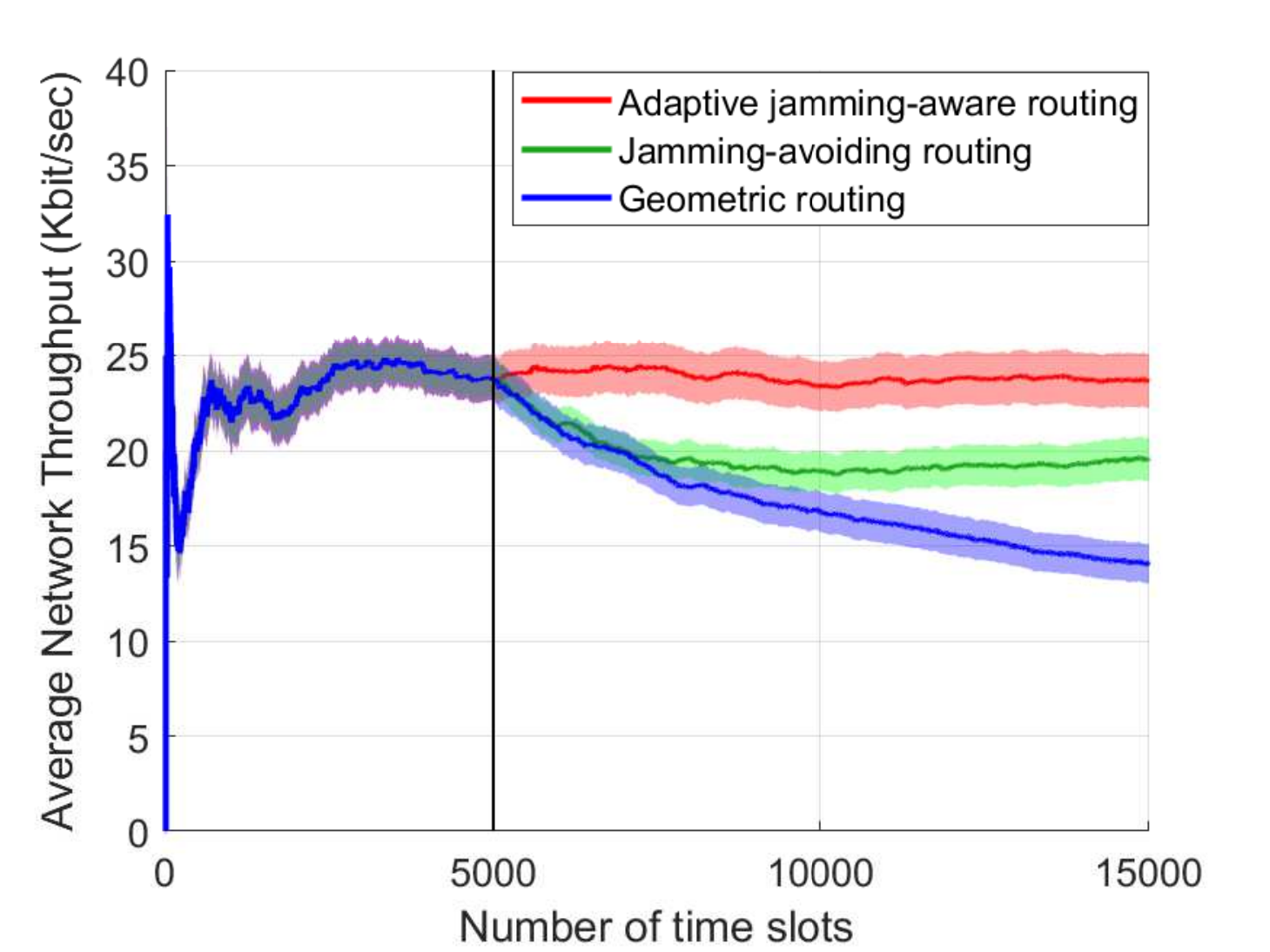}
	\caption{Average network throughput versus time for the three routing algorithms.}\label{throughputconf}
	}
\end{figure}

We measured the following metrics, the average end-to-end throughput, the proportion of failed eavesdropping  attempts, and the proportion of jammed red-force nodes. 
Figure \ref{throughputconf} shows the average network throughput over time. The shaded areas represent the $99\%$ confidence intervals of the average throughput of each of the studied protocols. As shown in Figure \ref{throughputconf}, the average throughput of the jamming-avoiding routing protocol and the adaptive, jamming-aware routing protocols does not reduce with time. Thus, the jamming-avoiding routing protocol and the adaptive, jamming-aware routing protocols are robust to red-force jamming. Figure \ref{throughputconf} also shows that, for the considered simulation scenario, adaptive, jamming-avoiding protocol achieves the highest average throughput. 

\begin{figure}[t]{
	\centering
	\includegraphics[width=7 cm,height=4 cm,angle=0]{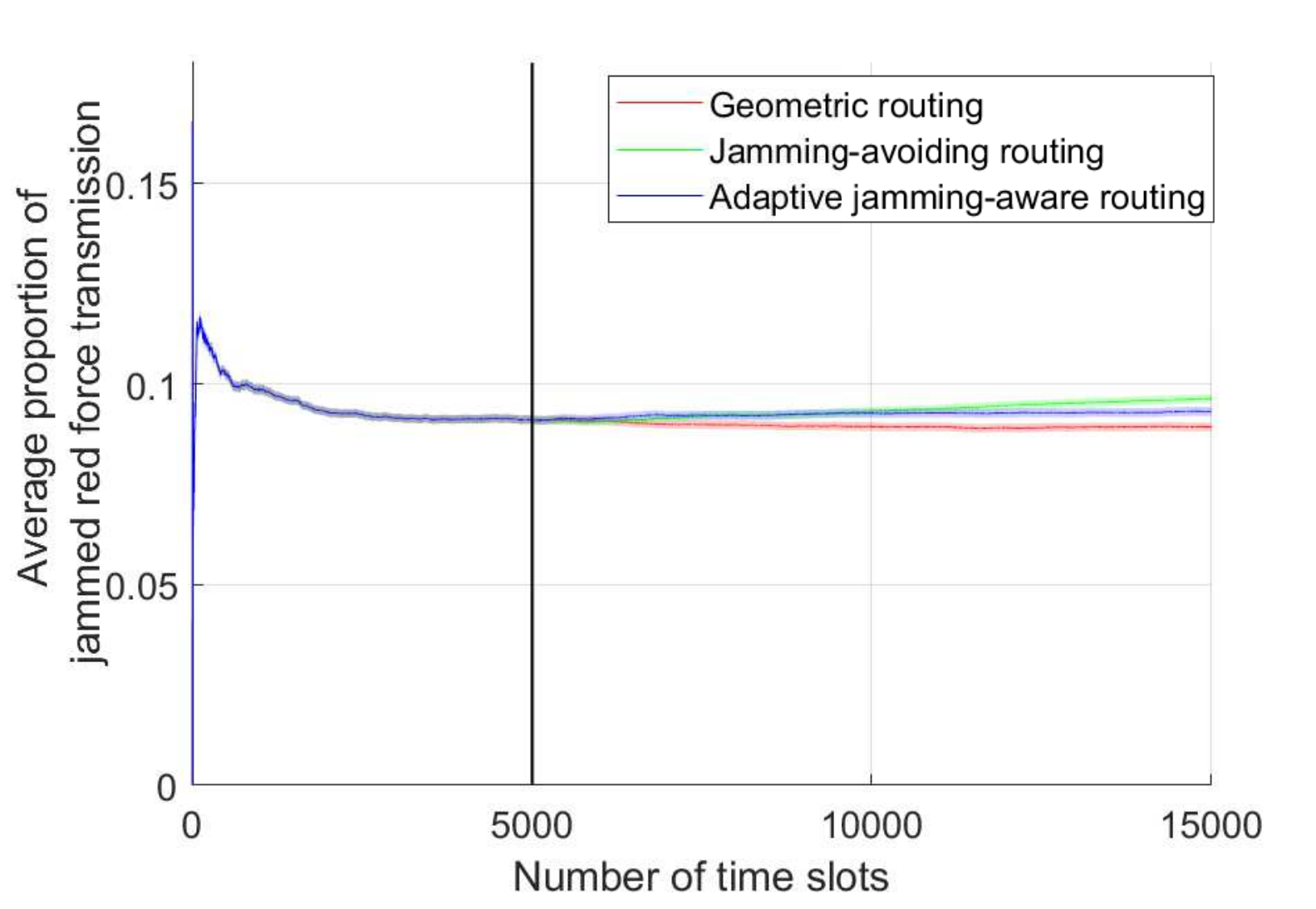}
	\caption{Average proportion of jammed red-force nodes versus time for the three routing algorithms.}\label{jamconf}
	}
\end{figure}

Figure \ref{jamconf} shows the average proportion of jammed red-force nodes over time. When red-force jamming starts, the jamming-avoiding routing protocol has higher proportion of jammed red-force nodes than the conventional minimum distance routing protocol does, whereas proportion of jammed red-force nodes using adaptive, jamming-aware routing is only $5\%$ compared to the conventional geometric routing, which shows that jamming/interference-aware routing protocols improve throughput while not significantly deteriorating  security.

\begin{figure}[t]{
	\centering
	\includegraphics[width=7 cm,height=4 cm,angle=0]{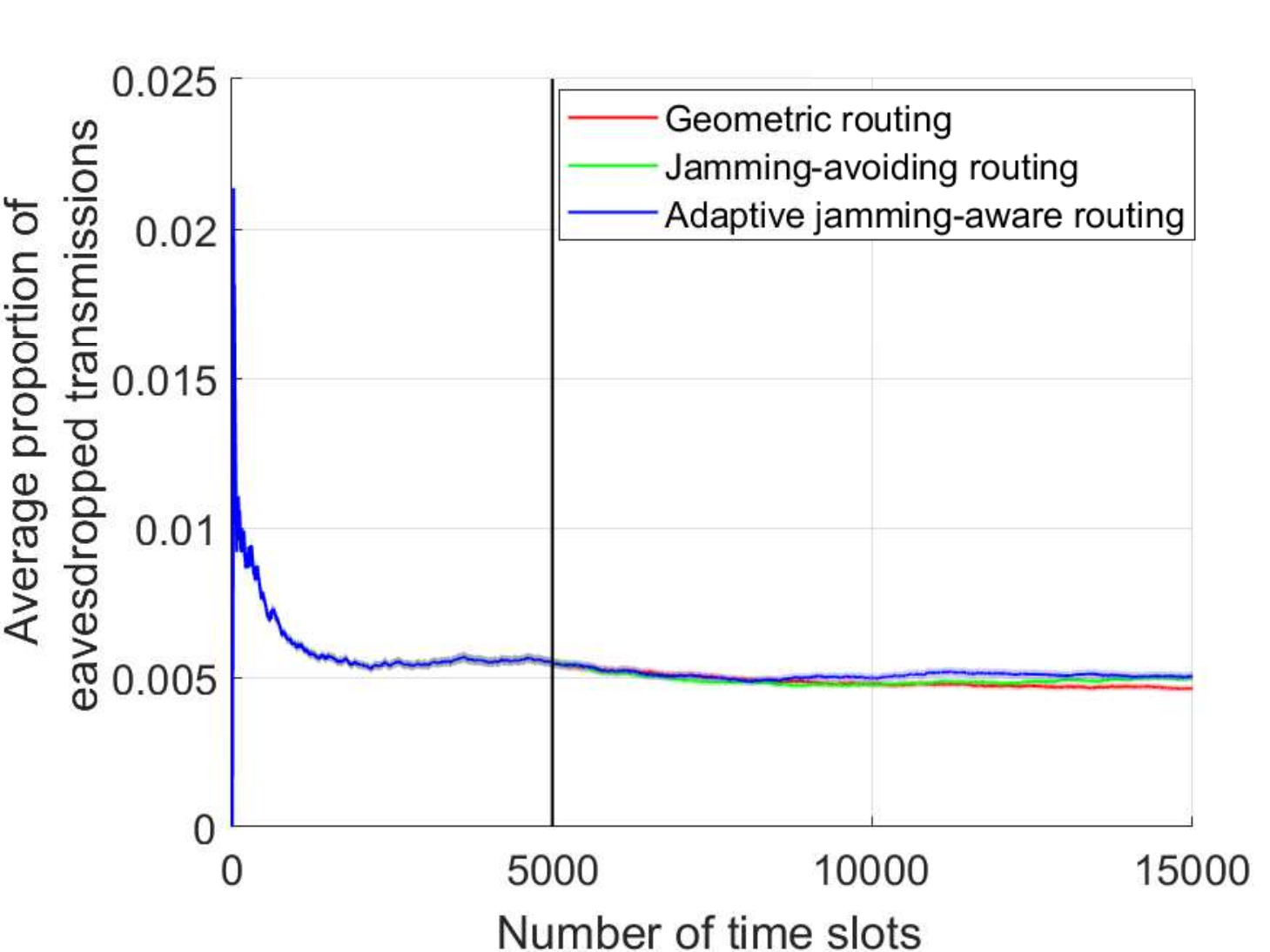}
	\caption{Average proportion of failed eavesdropping attempts versus time for the three routing algorithms.}\label{eavesconf}
	}
\end{figure}

 Figure \ref{eavesconf} shows the average proportion of failed eavesdropping attempts over time for the three considered routing protocols. When red-force jamming is initiated, the jamming-avoiding routing protocol has higher proportion of failed eavesdropping attempts than the conventional minimum distance routing protocol does, whereas the proportion of failed eavesdropping attempts using adaptive, jamming-aware routing protocol is only $4\%$ lower than the conventional geometric protocol, which shows that it improves throughput without jeopardizing security.

\begin{figure}[t]{
	\centering
	\includegraphics[width=7 cm,height=4 cm,angle=0]{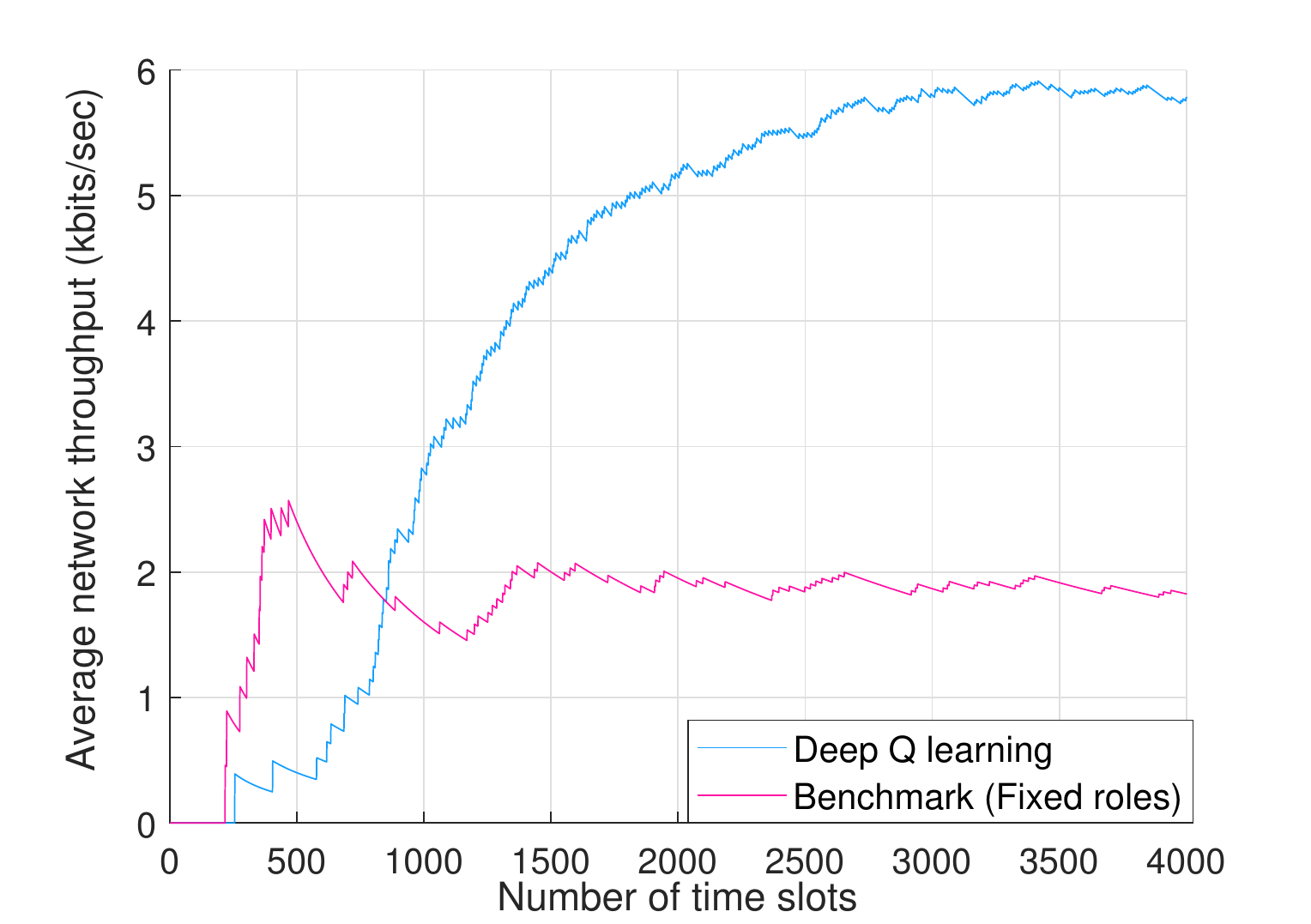}
	\caption{Average throughput versus time using deep Q-learning.}\label{deepthro}
	}
\end{figure}

\begin{figure}[t]{
	\centering
	\includegraphics[width=7 cm,height=4 cm,angle=0]{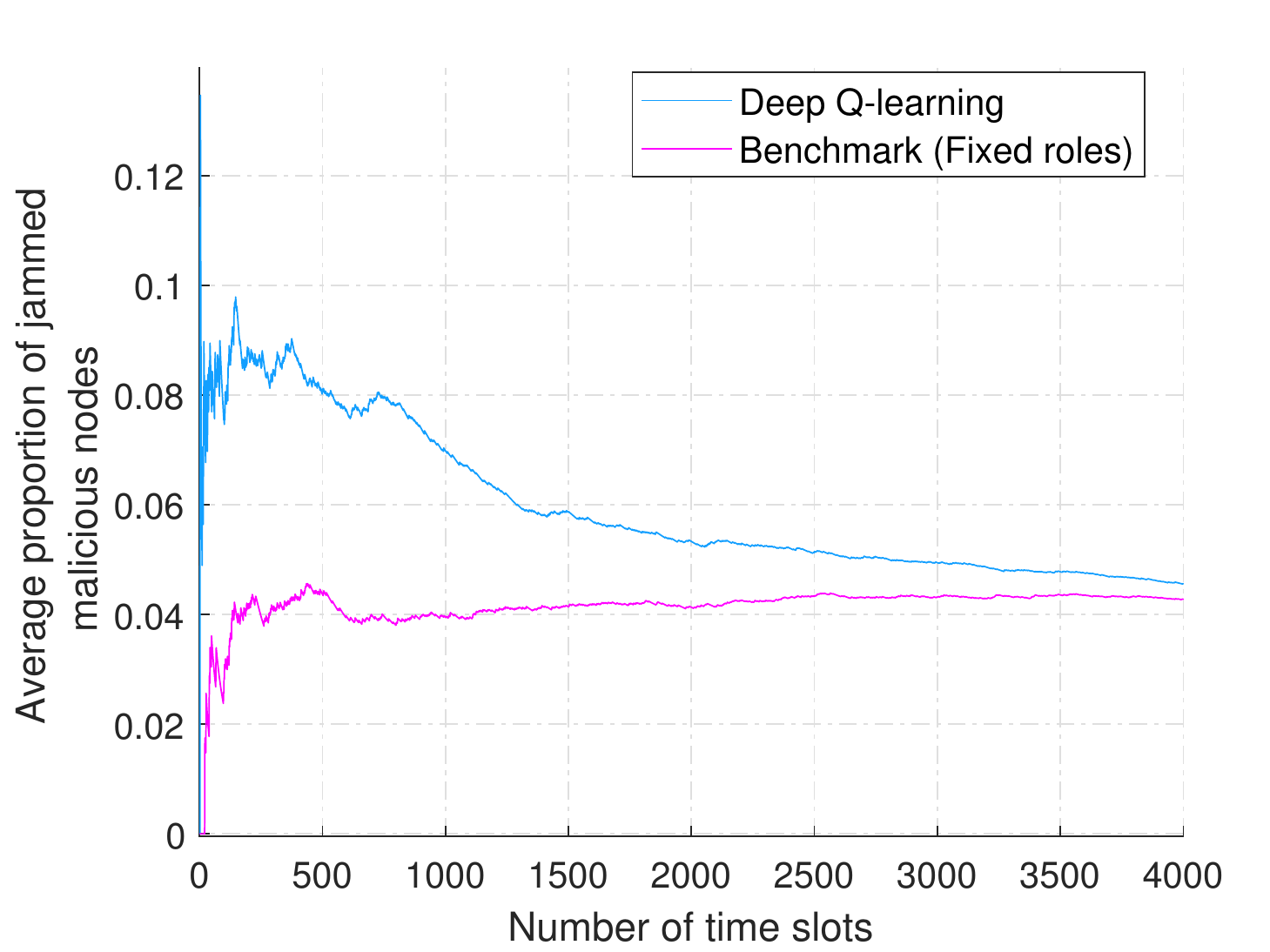}
	\caption{Average proportion of jammed red-force nodes versus time using deep Q-learning.}\label{deepjam}
	}
\end{figure}

\begin{figure}[t]{
	\centering
	\includegraphics[width=7 cm,height=4 cm,angle=0]{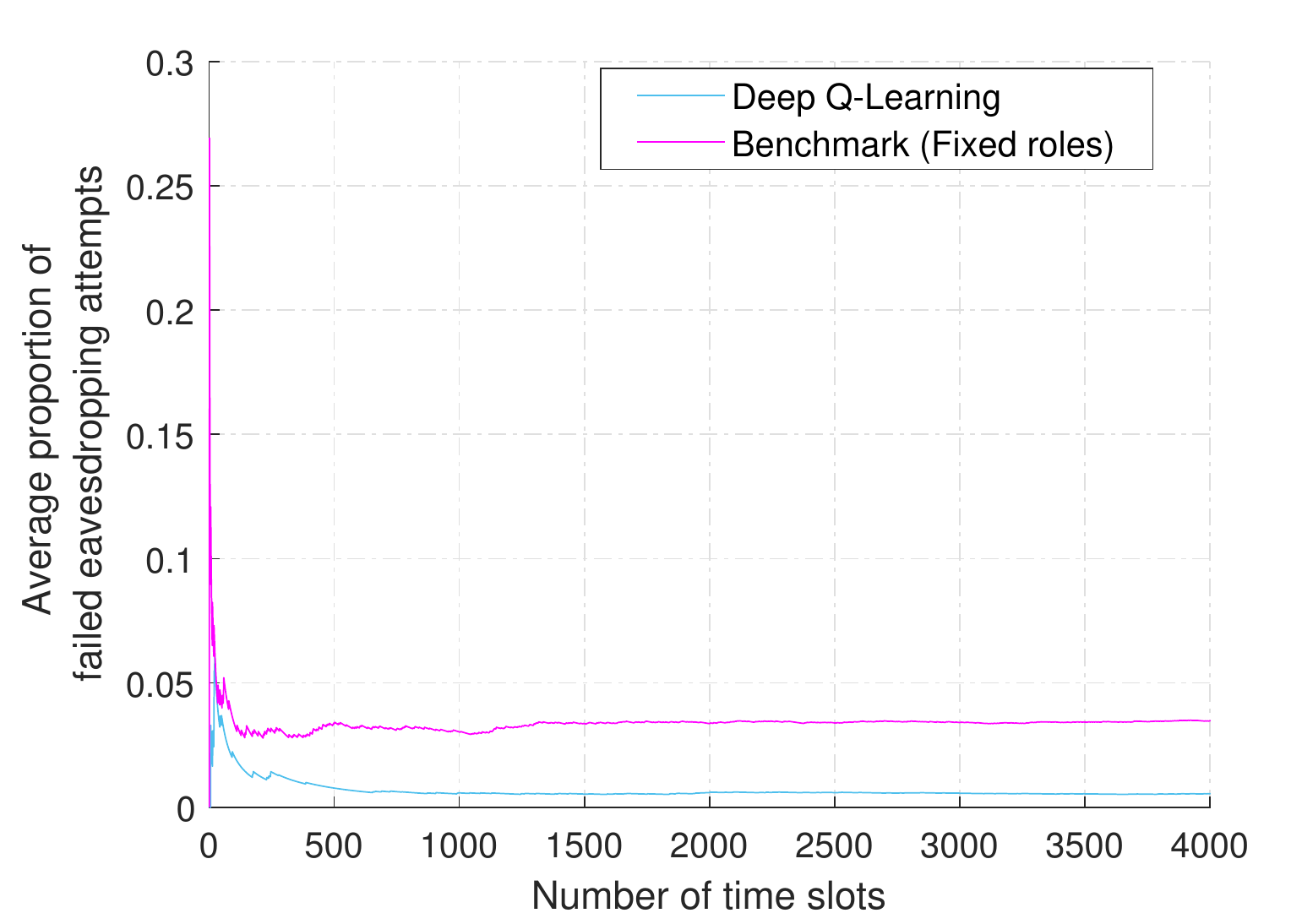}
	\caption{Average proportion of failed eavesdropping attempts versus time using deep Q-learning.}\label{deepeaves}
	}
\end{figure}

\subsection{Deep reinforcement learning results}
Adaptive, jamming-aware geometric routing is used as the routing protocol. The actor and critic networks are developed as feedforward neural networks (FNNs), each with two hidden layers. The ReLU function is used as the activation function. We set the episode size to be $8$, the batch size to be $2$, and the memory capacity $C$ to be $10000$. 
For better performance, we transform the input into a one-hot input vector. As for exploration, we set the value of epsilon to be $1$ for the first $500$ time slots, and then to $0.2$ for the next $300$ time slots, and, afterwards, it is set to $0.01$ until the end of the simulations. 

The weights of the immediate payoffs are $w_T = 15$, $w_{AJ}= 3$, $w_{CJ}= 5$, $w_E = 3$, and $w_D=1$ (namely, throughput is prioritized). We compute the average throughput achieved using the deep Q-learning algorithm and the baseline in which nodes have fixed communication or security roles. Figure \ref{deepthro} shows the average throughput achieved for the considered cases over time. When the deep Q-learning algorithm is used, the average throughput increases with time until it stabilizes around $3,000$ time slots. As shown in Figure \ref{deepthro}, the proposed deep Q-learning approach can achieve considerable gain, in terms of throughput, compared to the case when the communication and red-force nodes assume fixed roles.

Figure \ref{deepjam} shows the average proportion of jammed red-force nodes over time. Using Q-learning, the average proportion of jammed red-force nodes decreases with time because weighing factors are set to prioritize throughput, which causes the average proportion of jammed red-force nodes to decrease as nodes learn correctly their roles using deep Q-learning. Similarly, Figure \ref{deepeaves} shows that the average proportion of failed eavesdropping attempts decreases with time using deep Q-learning, as nodes are gradually learning to optimize the throughput.

\section{Conclusion} \label{Sec:Sec5}
In this paper, we designed a deep reinforcement learning approach that enables nodes to dynamically select their roles to maximize the network performance while ensuring robustness against jamming and eavesdropping attacks. Simulation results demonstrate the robustness of the proposed jamming-aware routing protocol and show that when throughput is prioritized, the proposed deep reinforcement learning approach achieves a three-fold increase in throughput, compared to a benchmark policy that assumes that the nodes have prior fixed roles.


\begin{thebibliography}{00}

\bibitem{secQoS1}Y. Xu, J. Liu, O. Takahashi, N. Shiratori and X. Jiang, ``SOQR: Secure optimal QoS routing in wireless ad hoc networks,'' \emph{IEEE Wireless Communications and Networking Conference (WCNC)}, 2017.

\bibitem{secQoS2} M. Hashem Eiza, T. Owens and Q. Ni, ``Secure and robust multi-constrained QoS Aware routing algorithm for VANETs,'' \emph{IEEE Transactions on Dependable and Secure Computing}, 2016.

\bibitem{RL1}J. Boyan, M.L. Littman, ``Packet routing in dynamically changing networks: A reinforcement learning approach,'' \emph{Advances In Neural Information Processing Systems}, 1994.

\bibitem{RL2} R. Sun, S. Tatsumi, and G. Zhao, Q-map,`` A novel multicast routing method in wireless ad hoc networks with multiagent reinforcement learning,''  \emph{IEEE Conference on Computation, Communication, Control and Power Engineering}, 2002.

\bibitem{RL3} Y. Chang, T. Ho and L. P. Kaelbling, ``Mobilized ad-hoc networks: a reinforcement learning approach,'' \emph{International Conference on Autonomic Computing}, 2004.

\bibitem{RL4} W. Usaha, J.A. Barria,  ``A reinforcement learning ticket-based probing path discovery scheme for MANETs,'' \emph{Ad Hoc Networks},  2004.

\bibitem{Arul}
K. Arulkumaran, M. P. Deisenroth, M. Brundage, and A. Bharath, ``Deep reinforcement learning, A brief Survey," \emph{IEEE Signal Processing Magazine}, 2017.

\bibitem{Brownian}R. Groenevelt, E. Altman, and P. Nain, ``Relaying in mobile ad hoc networks: The Brownian motion mobility model'' \emph{Wireless Networks},  2006.

\bibitem{DQN} V. Mnih et al., ``Playing Atari with deep reinforcement learning,'' \emph{NIPS Deep Learning}, 2013.

\bibitem{DRL1}G. Han, L. Xiao and H. V. Poor, "Two-dimensional anti-jamming communication based on deep reinforcement learning," \emph{IEEE International Conference on Acoustics, Speech and Signal Processing (ICASSP)}, 2017.

\bibitem{jamming}
Y. E. Sagduyu, R. Berry, and A. Ephremides, ``Jamming games in wireless networks with incomplete information," \emph{IEEE Communications Magazine}, 2011.

\bibitem{Erpek}
T. Erpek, Y. E. Sagduyu, and Y. Shi, ``Deep learning for launching and mitigating wireless jamming attacks," \emph{IEEE Transactions on Cognitive Communications and Networking}, 2019.

\bibitem{Davaslioglu}
Y. Shi, K. Davaslioglu, Y. E. Sagduyu, W. C. Headley, M. Fowler, and G. Green, ``Deep Learning for Signal Classification in Unknown and Dynamic Spectrum Environments," \emph{IEEE International Symposium on Dynamic Spectrum Access Networks (DySPAN)}, 2019.

\bibitem{DRL2}Y. Li, X. Wang, D. Liu, Q. Guo, X. Liu, J. Zhang,  Y. Xu, ``On the performance of deep reinforcement learning-based anti-jamming method confronting intelligent jammer''.\emph{Applied Sciences},  2019.

\bibitem{comm_EW}
Y. E. Sagduyu, S. Soltani, T. Erpek, Y. Shi and J. Li, ``A unified solution to cognitive radio programming, test and evaluation for tactical communications," \emph{IEEE Communications Magazine}, 2017.


\bibitem{Yener} E. Tekin, A. Yener, ``The Gaussian multiple access wire-tap channel: Wireless secrecy and cooperative jamming,'' \emph{Information Theory and Applications Workshop}, 2007.


\end{thebibliography}
\end{document}